Furthering a Comprehensive SETI Bibliography

Julia LaFond

Department of Geosciences, The Pennsylvania State University, 303 Deike Building University Park, PA 16802, USA

jkl98@psu.edu

Jason T. Wright

Department of Astronomy & Astrophysics, Penn State Extraterrestrial Intelligence Center, Center for Exoplanets and Habitable Worlds, The Pennsylvania State University, 525 Davey Laboratory, University Park, PA 16802, USA

Macy J Huston

Department of Astronomy & Astrophysics, Penn State Extraterrestrial Intelligence Center, Center for Exoplanets and Habitable Worlds, The Pennsylvania State University, 525 Davey Laboratory, University Park, PA 16802, USA

In 2019, Reyes & Wright used the NASA Astrophysics Data System (ADS) to initiate a comprehensive bibliography for SETI accessible to the public. Since then, updates to the library have been incomplete, partly due to the difficulty in managing the large number of false positive publications generated by searching ADS using simple search terms. In preparation for a recent update, the scope of the library was revised and reexamined. The scope now includes social sciences and commensal SETI. Results were curated based on five SETI keyword searches: "SETI", "technosignature", "Fermi Paradox," "Drake Equation", and "extraterrestrial intelligence." These keywords returned 553 publications that merited inclusion in the bibliography that were not previously present. A curated library of false positive results is now concurrently maintained to facilitate their exclusion from future searches. A search query and workflow was developed to capture nearly all SETI-related papers indexed by ADS while minimizing false positives. These tools will enable efficient, consistent updates of the SETI library by future curators, and could be adopted for other bibliography projects as well.

1. Introduction

In 2019, Alan Reyes and Jason Wright used the Astrophysics Data System (ADS) to create a living bibliography for the field of the Search for Extraterrestrial Intelligence (SETI) [1]. In their announcement, they reasoned that a public bibliography could: clarify the history of the field, provide context for the emergence and use of SETI-specific terms, improve public perception of SETI by demonstrating the depth and breadth of scholarship on the topic, enable the quantification of searched parameter space, establish criteria for what constitutes a SETI paper, and improve attribution and citation within the field.

Furthermore, they argued that SETI is a small enough field that the bibliography could grow to become comprehensive.

This bibliography may be accessed at ADS with the search query element "bibgroup:SETI," or by visiting https://ui.adsabs.harvard.edu/search/p_=0&q=bibgroup%3ASETI&sort=date%20desc%2C%20bibcode%20desc. This can be combined with other elements, for instance " author:"Wright, Jason" and bibgroup:SETI" produces all list of all SETI papers in ADS written by Jason Wright. Because ADS aims to be a comprehensive database of all astronomy-related publications, this should allow for efficient searching of the SETI corpus in the physical sciences, provided the library can be kept up to date and curated.

Unfortunately, updates to the bibliography have been incomplete; community participation in its maintenance has been low, and there is no designated curator or network of curators to maintain the bibliography. Furthermore, the sheer volume of papers contained on ADS complicates maintenance: a search for the term "SETI", for instance, returns over 2,000 results. It would be time-consuming for anyone to sift through the database for newly released papers, let alone for older publications that have since been added to ADS or were missed by previous searches. Automation would be difficult for two reasons: false positives and subjectivity. False positives result when keyword searches identify non-SETI publications, such as papers about the Seti River Basin in Nepal. Subjectivity results from the inherent grey areas of classification; for example, papers about the Drake Equation that deal with non-intelligent life are generally considered astrobiology, but not SETI.

Curation of the library by an individual or team could ensure consistent updates of the library. However, future curators need a standard procedure both to ensure consistency and to reduce the time and effort needed to maintain the library. Therefore, updating the library provided an opportunity to set a precedent for future curation procedures.

The goals of the project were as follows:

Update the SETI library.
Locate all SETI publications contained in ADS that are not already in the SETI library. Add them to the SETI library.
Refine criteria for publication inclusion.
Provide clearer definitions of which results should be added to the SETI library. Increase the scope where applicable.
Create a search method that locates a high number of SETI papers while returning a low number of false positives.
Develop a procedure for efficient library updates.

2. Revised Criteria
The creators of the library outlined several criteria for the inclusion of publications [1]. Acceptable topics included: SETI observation papers (including artifact SETI and null results), SETI-specific instrumentation, The Fermi Paradox, Meta-SETI (papers about the field as a whole), and anything fundamental to the history of SETI. Allowable formats were restricted to papers, books, and abstracts. They also defined four topics as out-of-bounds for the bibliography: pure astrobiology, fundamental physics and engineering, social sciences, and pseudoscience.

Although the authors agreed social science is relevant to SETI, it was excluded based on its poor representation in ADS [1]. Recent follow-up searches revealed that ADS contains many social science papers focused on SETI. Though this collection is surely a small fraction of the totality of social science SETI work, there was a high enough volume of publications to justify expansion the scope of the bibliography to include them. This means that the bibliography's incompleteness to social science papers on SETI will match that of ADS.

The exclusion of pseudoscience, though well-merited, could be difficult for future curators. Frequently, only a title and abstract are available for perusal, and even if the full text is available, reviewing the full contents of every paper would considerably slow cataloging efforts. Furthermore, it is the only topic in which the curator would need to verify the quality of the publications, rather than their relevance to SETI. Finally, the category is left undefined, with no criteria for determining the boundary between science and pseudoscience. For the sake of consistency, this criterion now only excludes entries that are openly satirical or are primarily concerned with UFOs and alien abductions as phenomena caused by extraterrestrial intelligences (ETI).

One topic not addressed by the original paper was commensal SETI (i.e., SETI performed simultaneously with studies of natural phenomena, sometimes more inaccurately called "parasitic" SETI). Due to the high volume of SETI research that is done commensally, it merits inclusion in the bibliography. Unfortunately for categorization efforts, commensal SETI tends to be entangled with fundamental physics/astronomy/engineering efforts. To determine whether a publication should be included, the following questions can serve as a litmus test: Is SETI one of the main goals of the mission, or else is the mission intentionally structured to enable commensal SETI? For engineering, is SETI one of the intended uses of the technology? Do the authors specifically discuss applications or significance to SETI beyond a single sentence in the introduction or discussion sections? Do the authors identify a target or method for future SETI searches? Is the work done necessary for future SETI missions or research, and do the authors acknowledge this? If the answer to any of these questions is "yes", the publication merits inclusion within the bibliography.

One other aspect to consider was the restriction of formats. The original intent to exclude non-academic work from the bibliography is correct. Excluding anything other than books, abstracts, and papers, however, may exclude important SETI sources. Some publications, such as government documents or software, are still relevant to SETI as a field, and should be included in a comprehensive bibliography. Therefore, only book reviews should be automatically excluded.

3. Workflow

To locate a population of publications to begin the update, libraries were created based on the following search terms: "SETI", "technosignature," "Drake Equation," "Fermi Paradox," and "Extraterrestrial Intelligence." ADS allows one to perform set operations between libraries, allowing users to take the difference of the search-term libraries and the SETI library to examine only entries not yet included.

There are several common sources of false positives. For "SETI," there is the Seti River Basin in Nepal, the physics researcher Julia Seti, and non-SETI research by the SETI institute. The term "technosignature" is robust, but does not include results for its plural, "technosignatures." The term "Drake Equation" frequently finds papers by various Drs. Drake, as well as astrobiological papers that did not discuss ETI. "Fermi Paradox" also returns results for the "Fermi-Pasta-Ulam-Tsingou Paradox," a completely unrelated phenomenon. Finally, ADS automatically searches for acronyms of search terms, so "Extraterrestrial Intelligence" returns all instances of "ETI" unless an exact search is used, which might eliminate hits for, for instance, "Extra-terrestrial intelligence." Unfortunately, several fields use the same acronym to refer to different things: temperature indices or indicators, topological insulators, elastic tensor imaging, effector-triggered immunity, energy-tracking impulse, energy-technology installation, electrothermal instability, and electronic-transport-informatics, to name a few.

Confusion from authors can be eliminated by restricting the search to abstracts and keywords (abs:) or the body of the paper (body:). Restricting broad search terms to abstract and keyword searches, rather than full-text searches, can filter out papers that only make passing mention of the subjects or of confounding false positive terms. The downside of this approach is that some papers that merit inclusion in the library only make passing mention of the selected keywords. The use of abstract or full-text searches becomes a trade-off between false positives and false negatives, and optimization depends on the keyword in question.

Specific keywords that cause confusion can be excluded from the search using the "NOT" qualifier. The "AND" operator can be used to return only publications that meet multiple conditions; for instance, narrowing down Drake Equation papers to ones that deal with intelligent life. The "OR" operator allows multiple keywords to be searched simultaneously.

The "Not SETI" library was created from publications that did not meet the requirements for inclusion to screen them out of future searches. Using the ADS set operators, it is possible to add the search term of the form "NOT docs(library/qazeXzDlSj-d06qbiWLoXQ)" to exclude them from all future searches (where the string of characters is a key assigned by ADS to that library). The SETI library can be excluded the same way.

Based on trial and error, a good formulation for future searches of ADS for new entries is:

body:("Fermi Paradox" NOT "Pasta") OR abs:("SETI" NOT "Nepal") OR body:("Drake Equation" AND "intelligence") OR body:("technosignature") OR body:("technosignatures") OR abs:("Extraterrestrial Intelligence" NOT "Elastic Tensor Imaging" NOT "Exceptional Topological Insulator" NOT "Temperature Index" NOT "Temperature Indicator" NOT "effector-triggered immunity" NOT "Energy-tracking Impulse" NOT "energy-technology installation" NOT "Electronic-Transport-Informatics" NOT "electrothermal instability") NOT docs(library/qazeXzDlSj-d06qbiWLoXQ) NOT docs(library/k1BwfM56QgKbl6X-PXADqg)

Therefore, the procedure for updating the SETI library is as follows (see also Fig. 1):

1. Run the search string.
2. Classify the results and add irrelevant publications to "Not SETI."

3. Add the relevant search results to "SETI."
4. Re-run the search string to ensure it returns no results. Repeat the process if results are found.

It is possible to restrict results by publication year (e.g. year: 2021). Current-year searches should be completed about once a month to prevent results from building up to overwhelming levels. However, curators should occasionally run the search without restricting the publication year to find any publications that either have been overlooked or were added to ADS after their publication year. Such a search should be completed every 6-12 months.

4. Addendum Demographics
On December 8, 2020, 553 entries were added to the SETI library.

431 of the addendum entries are articles, 171 of which were refereed. 122 entries are other publication types, including abstracts, books, proceedings, technical reports, press releases, PhD theses, software, and catalogs. 484 of the publications are from the astronomy collection, 138 are from the physics collections, and 9 are from the general collection.

As seen in Fig. 2, publications tend to skew toward recent years, with a large spike in 2018 and 2019. However, there are entries as far back as the 1960s, indicating a significant number of entries that were either overlooked, added after the library was created, or excluded under previous library curation criteria.

Fig. 3 provides citation statistics for the addendum. The addendum received 1345 total citations. The number of citing papers, or the number of unique papers that cite the contents of the addendum, is 1077. The average citation rate for all the publications was 2.4, while the average citation rate for only refereed publications was 5.7.

The SETI bibliography has continued to be updated since the creation of the addendum. In consultation with James Davenport, monthly updates have been sent to SETI.news as part of the SETI.news mailing list. Anyone interested in receiving these mailers can sign up at SETI.news.

5. Discussion & Future Work
The project goals were both to update the SETI library and to develop a procedure to enable consistent and efficient updates. The creation of a search string and the "Not SETI" library will facilitate future updates. However, in ruling out false positives, the search string could inadvertently rule out publications relevant to SETI. There are two possible ways to circumvent this problem. One is to occasionally broaden search criteria to increase the influx of results. Unfortunately, the offset of this approach is time spent curating the results. The other is to echo the call of Reyes & Wright for members of the SETI community to send overlooked entries to the library maintainers or to SETI.news [1].

Recently, it has become possible to have multiple users edit a single ADS library, thanks to a recent feature update at ADS. Curators are now taking advantage of this feature, and as of January 2021, library curation is being managed primarily by Macy Huston; publications may be recommended to her at mhuston@psu.edu. A simplified search string is currently being used, which picks up more false positives but may reduce the chance of false negatives:

(body:("Fermi Paradox") OR body:("SETI") OR body:("Drake Equation") OR body:("technosignature") OR body:("technosignatures") OR body:("Extraterrestrial Intelligence") ) NOT docs(library/k1BwfM56QgKbl6X-PXADqg) NOT docs(library/qazeXzDlSj-d06qbiWLoXQ)

At the end of each month, vetted results from this search are added to a "This Month in SETI" library, which is used to seed the newsletter at SETI.news, as well as to the full SETI library. Entries from the past month that do not fit the criteria are added to the Not SETI library. As of February 2021, ADS returns over 5,000 search results to this string which were published before the year 2021. These will be sorted gradually into the appropriate SETI or Not SETI library, a subset at a time during each month's library update process.

## 6. Acknowledgements

This paper grew out of a final project in the 2020 graduate course in SETI at Penn State. We thank James Davenport for collaborating on an effort to use a modified curation procedure to generate results for and revive SETI.news and the associated newsletter. This research has made use of NASA's Astrophysics Data System, and we thank its maintainers for their assistance in the project, including giving us early access to the group editing features for libraries. The Penn State Extraterrestrial Intelligence Center and the Center for Exoplanets and Habitable Worlds are supported by the Pennsylvania State University and the Eberly College of Science.

## 8. Figures & Tables

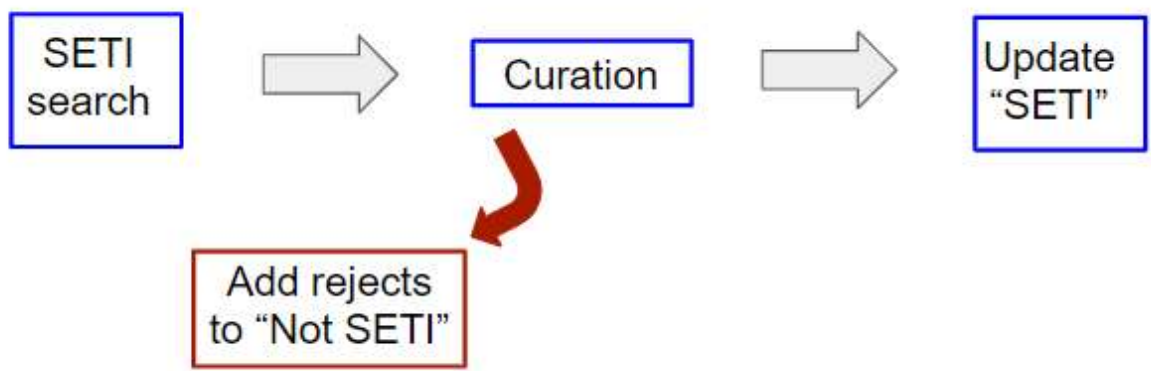

Figure 1: Example workflow for updating the SETI library: begin with the search string, curate, and update "Not SETI" and "SETI."

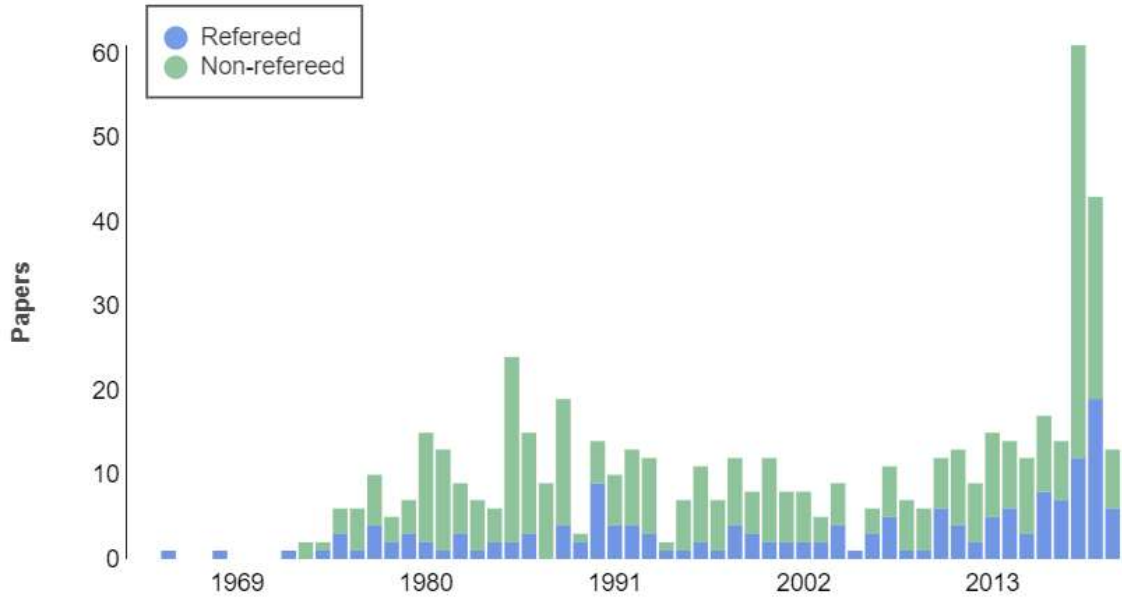

Figure 2: Distribution by publication year of the 553 publications contained in the December 2020 addendum to the SETI ADS library. Plot generated by ADS.

| | | Totals | Refereed |
|---|---|---|---|
| Number of citing papers | ❓ | 1071 | 800 |
| Total citations | ❓ | 1329 | 962 |
| Number of self-citations | ❓ | 67 | 45 |
| Average citations | ❓ | 2.4 | 5.7 |
| Median citations | ❓ | 0 | 1 |
| Normalized citations | ❓ | 700.5 | 444.9 |
| Refereed citations | ❓ | 1015 | 783 |
| Average refereed citations | ❓ | 1.8 | 4.6 |
| Median refereed citations | ❓ | 0 | 1 |
| Normalized refereed citations | ❓ | 526.3 | 360.6 |

Figure 3: ADS generated citation data of the 553 publications contained in the December 2020 addendum to the SETI ADS library. Plot generated by ADS.